# Logical Varieties in Normative Reasoning

**Mark Burgin**

Department of Computer Science
University of California, Los Angeles
Los Angeles, CA 90095, USA

**Kees (C.N.J.) de Vey Mestdagh**

Department of Legal Theory, Centre for Law&ICT
University of Groningen
The Netherlands

## ABSTRACT

Although conventional logical systems based on logical calculi have been successfully used in mathematics and beyond, they have definite limitations that restrict their application in many cases. For instance, the principal condition for any logical calculus is its consistency. At the same time, knowledge about large object domains (in science or in practice) is essentially inconsistent. Logical prevarieties and varieties were introduced to eliminate these limitations in a logically correct way. In this paper, the Logic of Reasonable Inferences is described. This logic has been applied successfully to model legal reasoning with inconsistent knowledge. It is demonstrated that this logic is a logical variety and properties of logical varieties related to legal reasoning are developed.

**Introduction**

Although each domain of knowledge is more or less affected by the problem of inconsistent knowledge, this issue is particularly intense in the domain of legal knowledge, since it consists of the rules and procedures used to describe and solve legal conflicts, which presupposes contradictory and hence inconsistent perspectives. Human processors of legal knowledge follow formal and informal problem-solving methods in order to reduce the number of legal perspectives and eventually to decide, temporally and within a specific context, on a common perspective. The formal methods are based on universal properties of formally valid legal argument. The informal methods are based on legal heuristics consisting in tentative legal decision principles. The first category can be formalized by logic because it applies peremptorily to all legal perspectives. The second category cannot be fully formalized by logic because, although it is commonly applicable, it can always be refuted by a contradictory decision principle and even by the mere existence of an underlying contradictory argument.

To eliminate these limitations in a logically correct way, the concepts of a logical prevariety and variety were introduced [3, 4]. Logical varieties represent the natural development of logical calculi and thus, they show the direction in which mathematical logic will inevitably go. They generalize in a natural way the concept of a logical calculus and are more advanced systems of logic. Logical varieties include logical calculi as the simplest case. There are different types and kinds of logical varieties and prevarieties:

*Deductive* or *syntactic varieties* and *prevarieties*.

*Functional* or *semantic varieties* and *prevarieties*.

*Model* or *pragmatic varieties* and *prevarieties*.

Semantic logical varieties and prevarieties are formed by separating those parts that represent definite semantic units. In contrast to syntactic and semantic varieties, model varieties are essentially formal structures.

Here we apply the theory of logical varieties to normative knowledge.

Constructive practical implications of the theory of logical varieties are:

1. Combination of different logical tools, e.g., different types of inference, and approaches/perspectives in one system.

2. Achieving multifuctionality of logic. The process is similar to trends in information technology: simple calculators do only calculations with numbers, while computers are multifunctional being able, for example, to work with texts and images; now cell phones become more and more multifunctional.

3. Provide efficient means to work with information that has been traditionally treated as inconsistent.

Cognitive practical implications of the theory of logical varieties:

1. Better and more precise description of reality.

2. Better and more precise understanding of reality.

To represent legal reasoning with formalized structures in efficient legal expert systems the Logic of Reasonable Inferences (LRI) has been developed and successfully applied in the legal knowledge domain [10 – 13].

In this paper, it is demonstrated that LRI is a logical variety and properties of logical varieties related to legal reasoning are studied.

**1. Logical varieties**

Syntactic varieties and prevarieties are built from logical calculi as buildings are built from blocks. That is why, we, at first remind the concept of a logical calculus.

Let us consider a logical language $L$ and is an inference language $R$.

**Definition 1.1.** A *syntactic* or *deductive logical calculus*, usually called *logical calculus*, is a triad (a named set) of the form

$$C = (A, H, T)$$

where $H \subseteq R$ and $A, T \subseteq L$, $A$ is the set of axioms, $H$ consists of inference rules (rules of deduction) by which from axioms the theorems of the calculus are deduced, and the set of theorems $T$ is obtained by applying algorithms/procedures/rules from $H$ to elements from $A$.

Let **K** be some class of syntactic logical calculi, $R$ be a set of inference rules, and **F** be a class of partial mappings from $L$ to $L$.

**Definition 1.2.** A named set (triad) **M** = $(A, H, M)$, where $A$ and $M$ are sets of expressions that belong to $L$ ($A$ consists of axioms and $M$ consists of theorems) and $H$ is a set of inference rules, which belong to the set $R$, is called:

(1) a *projective syntactic* (**K**,**F**)-*prevariety* if there exists a set of logical calculi $C_i = (A_i, H_i, T_i)$ from **K** ($A_i$ consists of axioms and $M_i$ consists of theorems of the logical calculus $C_i$) and a system of mappings $f_i : A_i \to L$ and $g_i : T_i \to L$ ($i \in I$) from **F** for which the equalities $A = \bigcup_{i \in I} f_i(A_i)$, $H = \bigcup_{i \in I} H_i$ and $M = \bigcup_{i \in I} g_i(T_i)$ are valid (it is possible that $C_i = C_j$ for some $i \neq j$).

(2) a *syntactic* **K**-*prevariety* if it is a projective syntactic (**K**,**F**)-prevariety where all mappings $f_i$ and $g_i$ that define **M** are inclusions, i.e., $A = \bigcup_{i \in I} A_i$ and $M = \bigcup_{i \in I} T_i$;

(3) a *projective syntactic* (**K**,**F**)-*variety* with the depth $k$ if it is a projective syntactic (**K**,**F**)-prevariety and for any $i_1, i_2, i_3, \ldots, i_k \in I$ either the intersections $\bigcap_{j=1}^{k} f_{ij}(A_{ij})$ and $\bigcap_{j=1}^{k} g_{ij}(T_{ij})$ are empty or there exists a calculus $C = (A, H, T)$ from **K** and projections $f: A \to \bigcap_{j=1}^{k} f_{ij}(A_{ij})$ and $g: T \to \bigcap_{j=1}^{k} g_{ij}(T_{ij})$ from **F**;

(4) a *syntactic* **K**-*variety* with the depth $k$ if it is a projective syntactic (**K**,**F**)-variety with depth $k$ in which all mappings $f_i$ and $g_i$ that define **M** are bijections on the sets $A_i$ and $T_i$, correspondingly.

(5) a (*full*) *projective syntactic* (**K**,**F**)-*variety* if for any $k > 0$, it is a projective syntactic (**K**,**F**)-variety with the depth $k$;

(6) a (*full*) *syntactic* **K**-*variety* if for any $k > 0$, it is a **K**-variety with the depth $k$.

We see that the collection of mappings $f_i$ and $g_i$ makes a unified system called a prevariety out of separate logical calculi $C_i$, while the collection of the intersections $\bigcap_{j=1}^{k} f_{ij}(A_{ij})$ and $\bigcap_{j=1}^{k} g_{ij}(T_{ij})$ makes a unified system called a variety out of separate logical calculi $C_i$. For instance, mappings $f_i$ and $g_i$ allow one to establish a correspondence between norms (laws) that were used in one country during different periods of time or between norms (laws) that are used in different countries.

The main goal of syntactic logical varieties is in presenting sets of formulas as a structured logical system using logical calculi, which have means for inference and other logical operations. Semantically, it allows one to describe a domain of interest, e.g., a database, knowledge of an individual or the text of a novel, by a syntactic logical variety and then to divide the domain in parts that allow representation by calculi.

In the case of projective syntactic (**K,F**)-prevarieties, the set *M* of logical formulas from a logical language *L* is represented by selecting a system of calculi $C_i$ from **K** and mapping theorems of these calculi into *L* so that all their images cover *M*. These calculi $C_i$ may have different languages $L_i$, different axioms (assumptions for reasoning) $A_i$ and/or different rules of inference $H_i$. However, all languages $L_i$ are amalgamated in *L* and all rules of inference $H_i$ are fused in *R* and represented in *H*. For instance, it is possible that $L = L_c \cup L_T \cup L_N$ where $L_c$ is the language of the classical predicate calculus, $L_T$ is the language of the tense logic, $L_N$ is the language of the logic of norms.

Besides, axioms (assumptions for reasoning) $A_i$ of the calculi $C_i$ represent the generating base (assumptions for reasoning) *A* of the syntactic variety **M** in a similar way.

Syntactic **K**-prevarieties have a better representation by calculi $C_i$ from **K**. Namely, they are unions of these calculi $C_i$ from **K**.

Projective syntactic (**K,F**)-varieties add one important feature to properties of projective syntactic (**K,F**)-prevarieties. Namely, not only components $C_i$ of the covering { $C_i$ ; $i \in I$} are calculi from **K** but also all intersections of the component images in *L* are presented by calculi $C_i$ from **K**.

Syntactic **K**-varieties properties of projective syntactic (**K,F**)-varieties and syntactic **K**-prevarieties. Namely, they are unions of these calculi $C_i$ from **K** and intersections of these calculi $C_i$ are also calculi from **K**.

**Definition 1.3.** The calculi $C_i$ used in the formation of the prevariety (variety) **M** are called *components* of **M**.

Inference in a logical variety **M** is restricted to inference in its components because at each step of inference, it is permissible to use only rules from one set $H_i$ applying these rules only to elements from the set $T_i$.

An interesting type of logical varieties was developed in artificial intelligence and large knowledge bases. As Amir and McIlraith write [1], there is growing interest in building large knowledge bases of everyday knowledge about the world, comprising tens or hundreds of thousands of assertions. However working with large knowledge bases, general-purpose reasoning engines tend to suffer from combinatorial explosion when they answer user's queries. A promising approach to grappling with this complexity is to structure the content into multiple domain- or task-specific partitions. These partitions generate a logical variety comprising the knowledge base content. For instance, a first-order predicate theory or a propositional theory is partitioned into tightly coupled subtheories according to the language of the axioms in the theory. This *partitioning* induces a graphical representation where a node represents a particular partition or subtheory and an arc represents the shared language between subtheories.

The technology of content partitioning allows reasoning engines to improve the efficiency of theorem proving in large knowledge bases by identifying and exploiting the implicit structure of the knowledge [1,8]. The basic approach is to convert a graphical representation of the problem into a tree-structured representation, where each node in the tree represents a tightly-connected subproblem, and the arcs represent the loose coupling between subproblems. To maximize the effectiveness of partition-based reasoning, the coupling between partitions is minimized, information being passed between nodes is reduced, and local inference within each partition is also minimized.

The tools and methodology of content partitioning and thus, implicitly of logical varieties are applied for the design of logical theories describing the domain of robot motion and interaction [1].

Concepts of logical varieties and prevarieties provide further formalization for local logics of Barwise and Seligman [2], many-worlds model of quantum reality of Everett [5, 6], and pluralistic quantum field theory of Smolin related to the many-worlds theory [9].

## 2. Reasoning in legal systems

Legal, or more broadly, normative knowledge is used to infer the normative characteristics of actual social situations. Legal knowledge is a subdivision of normative knowledge that is used in the formal legal (judicial) subsystem of social systems, such as countries, organizations or coalitions. Normative knowledge encompasses both normative opinions (know what) and the normative procedures (know how) that are used to infer these normative opinions. The normative characteristics that are inferred represent the mutual expectations of people about the conduct of others (rules of conduct). In a formal legal context, these expectations are commonly labeled as 'rights' (to the realization of conduct of others) and 'obligations' (of others to behave in agreement with the expectations).

Normative opinions range from informal to formal. On the informal side we find moral principles, social scripts, protocols, (technical) instructions, rules of thumb, rules of play etc. On the formal side we find legislation, legal principles, jurisprudence, policy rules etc. Normative opinions can be of a general (uninstantiated) and of a specific (instantiated) character. Normative procedures consist of (1) procedures to list all the normative opinions about a given situation that can be inferred from the given situation combined with the set of pre-existing normative opinions of the parties concerned and (2) procedures to reduce the number of normative opinions about the given situation to a (local and temporal) common opinion for (not necessarily *of*) the parties concerned. Both procedures involve *legal reasoning*. The second procedure also involves *legal decision-making*. Legal reasoning in the first class of procedures is concerned with the inference of normative opinions about the given situation. We will refer to this as "*the object level*". Legal reasoning in the second class of procedures is concerned with the inference of normative opinions about the reduction of normative opinions (e.g. "the judge is obliged to decide for a legally valid opinion"). We will refer to this as "*the meta level*". In the next paragraphs, we will discuss the properties of legal knowledge that should be taken into consideration in order to be able to develop a tenable computational model.

Legal or normative reasoning has no unique qualities compared to reasoning in other domains of knowledge. Normative characteristics of social situations are inferred by plain logical deduction from (agreed or disputed) facts and normative opinions. Normative opinions can have facts (e.g. the conduct of others) or opinions as their subject. In the former case, expectations about the conduct of others in general are inferred (the object level). In the latter case, expectations about the application of opinions are inferred (the meta level). To be precise: at both levels, expectations about the conduct of others are inferred. At the object level this relates to conduct in general, while at the meta level it relates to conduct concerning the application of normative opinions. Consequently, there is no formal difference between legal reasoning at the two levels.

One could think that an idiosyncrasy of legal reasoning may be found in the above addition "agreed or disputed", but disagreements about facts and opinionated qualifications are part of every domain of knowledge. However, the representation of disagreements about facts and conflicting opinions and the (local and temporal) resolution of these disagreements and conflicts is the aim of, and therefore essential to, the practical application of legal knowledge. What is special about this particular aim is the local and temporal character of the resolution. The aim of the application of legal knowledge in a social situation is to decide on a common perspective in order to be able to act in a coordinated manner. The decision does not (necessarily) cause facts or individual opinions to change; it simply introduces a new fact, that of the common perspective. It is even necessary for all the disputed facts and opinions to be represented permanently because they are not only part of the decision-making process but remain part of the legitimation of the common perspective.

The continued representation of disputed facts and opinions, even after a decision regarding a common perspective has been made, is not only essential to the legitimation of the decision. Legal knowledge is ultimately dynamic, meaning not only that people can change their opinion sequentially over time but that they can also hold different opinions in parallel at any given time. Normative opinions change and differ with time and given context. A common perspective only holds for the given situation of the parties concerned. Furthermore, the parties need not merely maintain their individual opinions in parallel with the common, decided opinion, but they may also immediately renounce the common opinion either individually or in unison. It is not uncommon that parties decide to act *contra legem*, for example to maintain the status quo or just to avoid a bagatelle.

The world is not transformed into a consistent state as a consequence of the completion of the legal proceedings. Agreement is reached within one context, at one moment, in order to complete a singular legal transaction (e.g. a verdict). The judge and all other parties can stick to their original opinions in every other transaction, but they may and frequently will also change their opinions in the aftermath of legal proceedings. People may also continue to act in violation of a verdict. A verdict may be overruled or be revised. And even the law may change.

The preceding description of legal knowledge and legal reasoning renders any normative opinion relatively *legally valid* (i.e. legally valid within its own context) and thus allows the existence of contradictory opinions. Fortunately, there are some universal constraints that reduce the number of opinions that can be taken into consideration. These constraints are based on the legitimation principle, which is universally acknowledged in legal disputes and which comprises amongst others the principles of legal justification and legal rationality. The principle of legal justification demands that each derived normative opinion is based on a complete argument, meaning that the opinion reached is supported by facts and grounded opinions. The principle of rationality comes down to the demand that the derived opinion and the argument it is based on are non-contradictory. Psychologically, these demands amount to common characteristics of human cognition. Formally, they boil down to the requirements of valid deduction and consistency. A logic modeling legal reasoning should abide by these requirements. The formal demands of valid deduction and consistency of opinions and their justifications reduce the number of *formally valid* opinions (reasonable inferences), but in most cases they do not enable a reduction to a single common opinion. Unfortunately, there are no further formal (absolute) criteria to reduce all the remaining alternative formally valid legal opinions to a single common opinion. Logical varieties can be used to model this fundamental inconsistency in a logically correct way.

### 3. The logic of reasonable inferences

The Logic of Reasonable Inferences (LRI) models legal reasoning, using the language of classical first-order predicate calculus, as this language seems powerful enough to express legal rules and factual situations without losing any relevant information [10–13].

The classical definition of semantic derivability seems to be a fairly reasonable one, but it enjoys a peculiar property if theories are allowed to be inconsistent: anything can then be derived from them! Thus, if $\Gamma$ is an inconsistent theory, then $\Gamma \vDash \varphi$ for any $\varphi \in \mathcal{L}$. Theories like this are called *trivial*, and logics that render inconsistent theories trivial are called *explosive*.

Explosiveness conflicts with any intuitive understanding of derivability. We surely do not want to conclude from an inconsistent theory on environmental law that the obligation to possess an environmental permit implies that one does not perform activities related to the environment, or that all farmers are civil servants. One is not liable to accept any derivation of a formula containing concepts not present in the theory from which it was derived.

Inconsistent theories, which model the body of rules of law, have their use in legal reasoning, as it is demonstrated in Section 2. Therefore, a relevant definition of semantic derivability must surely avoid the property of predicate calculus derivability concerning inconsistent theories by responding to inconsistent theories along the lines described above. This can be achieved by demanding that every justification for a derived conclusion is internally consistent, where a justification is the set of rules and observations (facts) used to derive the conclusion. This demand is a straightforward observation taken from legal reasoning theory.

These constraints lead to the definition of a new (non-explosive) semantic derivability relation $\vDash_r$ for the Logic of Reasonable Inferences (LRI). The language of the LRI is the language of the classical first-order predicate calculus. However, in a conventional logic (logical calculus), rules of inference depend only on the form of logical formulas. Thus, the LRI is not a conventional logic because its rules depend on *partitioning* of its language. We show that the LRI is a logical variety.

**Definition 3.1.** A *domain of rules*, or a *reasonable base*, in $\mathcal{L}$ is a tuple $\Delta$ defined as

$$\Delta = (A, H)$$

where $A$ and $H$ are sets of *wffs* in $\mathcal{L}$, such that $A$ is the consistent set of *axioms* for a reasonable theory, and $H$ is the set of (tentative) *assumptions* (hypotheses).

The *assumptions* model the rules of law that may or may not be applied in a given factual situation to derive a conclusion and contain all normative or subjective classifications of the factual situation. The *axioms* are intended to be valid in every justification and thus, restrict the number of possible justifications. These axioms represent the ascertained facts and previously ascertained conclusions (the permanent database in any implementation).

**Definition 3.2.** A *position* (or *conviction*) $\phi$ within a domain of rules $\Delta = (A, H)$ is a consistent set of wffs defined as

$$\phi = A \cup H'$$

where $H' \subseteq H$.

**Definition 3.3.** A position (conviction) $\phi$ within a domain of rules $\Delta = (A, H)$ is called *logically closed* if it is a predicate calculus.

Thus, a position is a set of rules taken from the domain of rules and represents a conviction of an individual or a group of people. Note that all positions should at least contain all axioms of the domain of rules and each position is consistent by definition. This shows that all logically closed positions form a logical variety in which all intersection are equal to $A$.

Let $\Delta$ be a domain of rules. Define a new semantic derivability-relation $\vDash_r$ as :

$$\Delta \vDash_r \varphi$$

*iff* there exists a position $\phi$ within $\Delta$ which satisfies

$$\phi \vDash \varphi$$

where $\vDash$ is the normal predicate calculus semantic derivability relation. If $\Delta \vDash_r \varphi$ holds, $\varphi$ is said to be a *reasonable inference* from the domain of rules $\Delta$.

This the exact form of inference in logical varieties.

We can paraphrase this definition by stating that a *wff* can reasonably be inferred from an inconsistent set of *wff* iff it is derivable (in the normal predicate calculus sense) from a consistent subset of this set which contains at least the axioms. Note that if a domain of rules $\Delta = (A, H)$ is consistent (i.e. if $A \cup H = \Gamma$ is consistent), then $\Delta \models_r \varphi \Box \Gamma \models \varphi$ behaves exactly like $\models$ when applied to consistent theories.

In this setting, a *justification* for a conclusion $\varphi$ derived from a domain of rules $\Delta$ is a minimal position (with respect to set-inclusion) $J$ within $\Delta$ such that $J \models \varphi$. This definition is based on the more intuitive definition as a set of rules and statements about the factual situation used to draw the conclusion. Note that a justification needs not be unique but it is always consistent, thus, satisfying our constraints.

A *context* in $\Delta$ is the union of $n$ simultaneously derived conclusions $\psi_i$ and their justifications $J_i$ derived from $\Delta$, i.e. a context is the set of tuples $\{ (\psi_i, J_i) \mid 1 \leq i \leq n \}$.

A context in $\Delta$ is called *consistent* if the justifications $J_i$ derived from $\Delta$ satisfy the following condition:

$$\text{The union } \bigcup_{i=1}^{n} J_i \text{ is consistent}$$

This guarantees that simultaneously derived conclusions are not based on mutually inconsistent positions, and that holds.

**Definition 3.4.** A *reasonable theory* Th $\Delta$ with a base $\Delta = (A, H)$ is the set of all wffs deducible in LRI from $\Delta$, i.e.,

$$\text{Th } \Delta = \{ \varphi \in \mathcal{L}; \text{ there is a position } \phi \in \Delta, \text{ such that } \phi \models \varphi \}$$

This shows that any reasonable theory Th $\Delta$ is a deductive logical variety $V$ of the form

$$V = \{ C_i ; i \in I \text{ and there is a position } \phi, \text{ such that } C_i = (\phi, d, T_\phi) \}$$

where $d$ is the set of all deduction rules of classical the first-order predicate calculus and $T_\phi$ is the set of all formulas deducible from the position $\phi$ by rules from $d$.

## 4. Compatibility in logical varieties

An important problem of logic is to combine logics and in particular, to include a system of calculi into one calculus. Gabbay [7] writes that "the problem of combining logics and systems is central for modern logic, both pure and applied. The need to combine logics starts both from applications and from within logic itself as a discipline. As logic is being used more and more to formalize field problems in philosophy, language, artificial intelligence, logic programming, and computer science, the kind of logics required becomes more and more complex." Logical varieties give a relevant context for solving this problem. Here we consider only deductive varieties.

Let **K** be a class of logical calculi and $M = \{C_i \,;\, i \in I\}$ be a deductive variety (**K**-variety).

**Definition 4.1.** A logical variety $M$ is called:

a) *Discrete* if its components are disjoint;

b) *Classical* if all its components are classical deductive calculi;

c) *Connected* if any two of its components have a non-void intersection;

d) *Compatible* (**K**-*compatible*) if it is a subset of a consistent calculus (of a calculus from **K**).

**Definition 4.2.** A set of components $\{C_i \,;\, i \in J\}$ of $M$ are called compatible if the subvariety of M generated by these components is compatible

**Lemma 4.1.** For any deductive variety **M**, there is a discrete deductive variety **DM** such that their upper levels are equal, i.e., $T(\mathbf{M}) = T(\mathbf{DM})$..

**Proposition 4.1.** If **M** is compatible (**K**-compatible), then **DM** is compatible (**K**-compatible).

**Theorem 4.1.** For any number $n > 1$ there is a classical connected deductive logical variety **M** with $n$ components such that any $n - 1$ components of **M** are compatible, but **M** is not compatible.

**Remark 4.1.** The condition that the variety is classical is essential.

## 5. Conclusion

Logical varieties and prevarieties eliminate certain limitations of conventional logical systems based on logical calculi, particularly, their explosiveness. The Logic of Reasonable Inferences is a logical variety and consequently has been successfully used to formally describe inconsistent normative knowledge and to develop legal decision support and expert systems.

Legal decision support systems and expert systems based on LRI, logical varieties and procedures of reasonable inference might be useful to judges, jurors, lawyers, detectives, and attorneys.

For instance, a Legal Decision Support System can help jurors and judges to find if witnesses are consistent in their depositions, if statements of different witnesses are compatible, if versions of persecution and defenders are consistent, and which of these versions is more grounded. A Legal Decision Support System can help a judge to find what laws and/or what precedence cases are more compatible with the given case. A Legal Decision Support System can help detectives and attorneys to find which conjectures are compatible with evidence and with one another and which of these conjectures are more grounded.